\documentclass{PoS}
\usepackage{wrapfig}
\usepackage{subcaption}
\usepackage{booktabs}
\usepackage{color}
\usepackage{multirow}
\usepackage[table]{xcolor}

\title{THE GEM QA PROTOCOL OF THE ALICE TPC UPGRADE PROJECT}

\ShortTitle{GEM QA protocol for the ALICE TPC upgrade}

\author{\speaker{Jens Erik Br\"ucken}\thanks{On behalf of the ALICE TPC upgrade collaboration.}\\
        Helsinki Institute of Physics, Helsinki, Finland\\
        E-mail: \email{erik.brucken@iki.fi}}

\author{Timo Hild\'en\\
        Helsinki Institute of Physics, Helsinki, Finland\\
        E-mail: \email{timo.hilden@helsinki.fi}}

\abstract{
The ALICE experiment at the Large Hadron Collider at CERN is upgrading its central tracking detector,  the Time Projection Chamber (TPC). The installation is foreseen during the second long shutdown of the Large Hadron Collider. The upgrade includes the complete exchange of the present MWPC readout chambers (ROC) with new ones based on Gas Electron Multiplier detectors. This is necessary due to the higher LHC luminosity and thus higher interaction rate. The new ROCs allow for continuous readout at 50~kHz compared to 500~Hz of the gated MWPC readout, while maintaining the particle identification capability of the present system. 

A thorough quality assurance scheme was developed to build a strict QA protocol.

The QA consists of two stages. The first stage, the basic QA is done close to the GEM production workshop at CERN and later at the framing and assembly centers. The second stage, the advanced QA is done at dedicated QA centers. Full traceability of detector components will be maintained throughout the process.  
A detailed description of the protocol will be given with emphasis on the high definition optical scanning and gain measurements of individual GEM foils. 

The production of the new ALICE TPC ROCs has finally started. First QA experience under production conditions and workload will be presented.
}

\FullConference{5th International Conference on Micro-Pattern Gas Detectors (MPGD2017)\\
		22-26 May, 2017\\
		Philadelphia, USA}

\begin{document}

\section{Introduction}

\noindent 
The Time Projection Chamber (TPC) of the ALICE experiment at CERN is currently upgraded and foreseen to be installed in the second large shutdown of the LHC. 
The new Readout Chambers (ROC) for the TPC are based on Gaseous Electron Multiplier technologies~\cite{GEM,mathis}. The R\&D and construction details can be found in the Technical Design Report~\cite{TPCTDR}. The baseline solution consists of a stack of four GEM foils each operated at a specific electric field configuration. This was developed to fulfill strict design criteria on energy resolution, ion back flow and operational stability. These design criteria can only be met by thorough Quality Assurance (QA) measures for the ROC production. 
The plan of the Quality Assurance (QA) procedures for the GEM foils has been introduced  already in~\cite{mpgd15,mball}.
The QA protocol has been further developed since. A refined description of the present QA protocol and first experiences of the production QA will be discussed.

The production of the ROCs has been successfully started during the first quarter of 2017. Also the GEM QA protocol has been finalized and is now in production mode. According to the plan at present time about 720 single GEM foils have to be tested. This number consists of 576 GEMs for the 36 ROCs with a total surface of $\sim128$~m$^2$ plus 25\% spares. The GEM foils come in four sizes.
Each ROC consists of a stack of four GEMS. In addition to standard foils, which have a pitch of 140~$\mu$m, two double pitch foils (280~$\mu$m) are introduced to optimize ion back flow.

\begin{wrapfigure}[18]{r}{0.39\textwidth}
\vspace{-0.2cm}\includegraphics[width=0.39\textwidth]{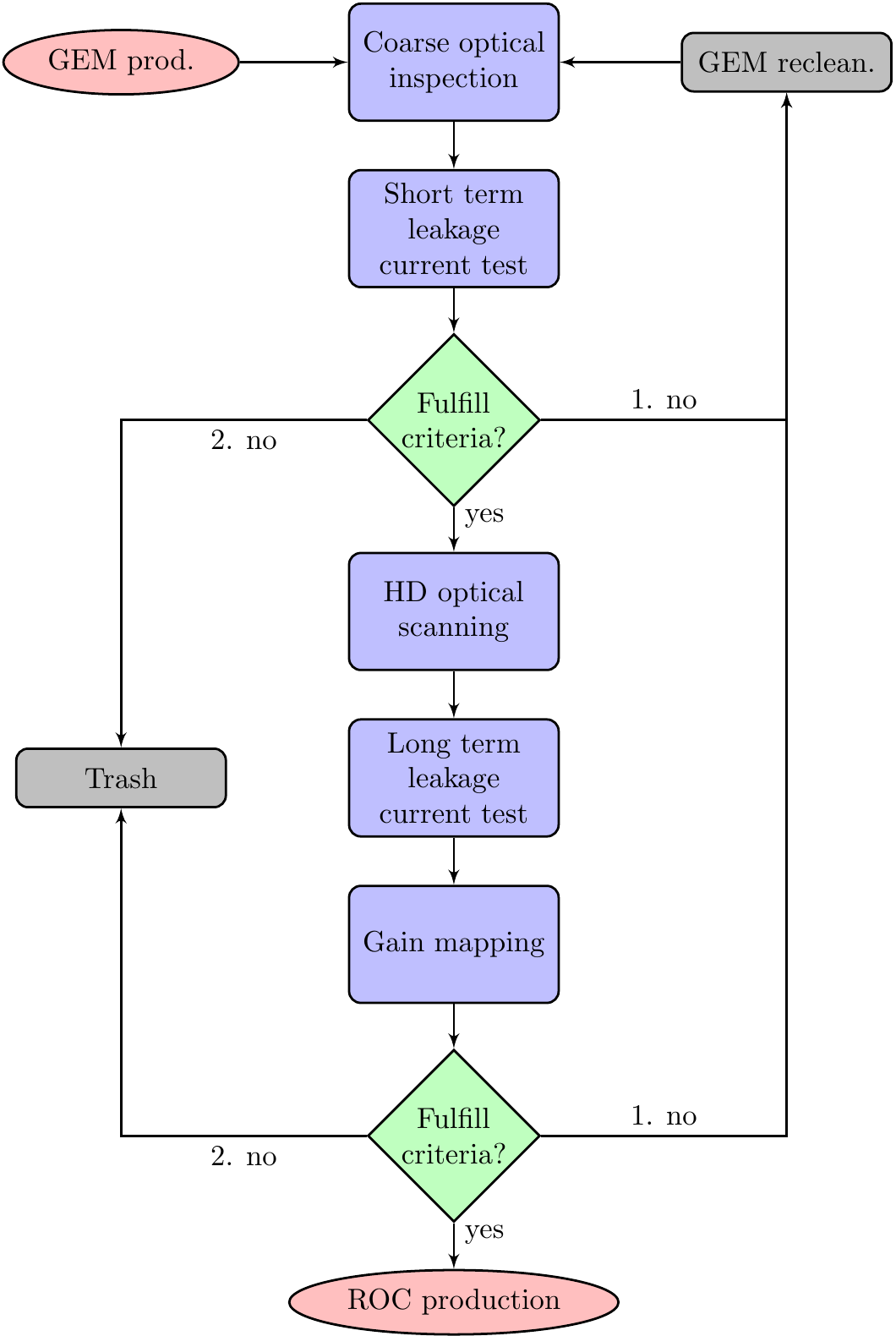}
\caption{Flowchart of the GEM QA production.\label{flowchart}}
\end{wrapfigure}

The GEMs that are produced with single-mask technique at the CERN PCB workshop\,\cite{ruiworkshop}, are transported in dedicated GEM Transport Systems (GTS) to the QA centres for testing. The premises for the QA centers are all clean-room facilities of at least class 1000 (ISO 6). GEMs are stored in dry cabinets. After all tests that are described here in detail are finished, GEMs are further transported to the framing- and ROC production sites.

\section{QA protocol}

\noindent The QA is performed in two stages. The first stage at CERN, called basic QA applies fast testing procedures to sort out foils at earliest possible stage. These procedures are also done after every transportation of the foils in particular in the four framing centers and the three assembly sites.

The second stage at the Wigner research center in Budapest, Hungary and the Helsinki Institute of Physics, Finland, called advanced QA applies more time consuming tests for qualitative certification. A flowchart of the QA production which will be explained in detail below is shown in Fig.\,\ref{flowchart}.

The results of the GEM QA tests will be entered into a common production database such that all relevant information is available on demand at later ROC production stages.

\subsection{Basic QA}
\noindent First the foils are inspected by eye for large defects $\gtrsim$1~mm (coarse optical inspection). Those can be over-etched- or under-etched holes, scratches, holes etched at the segment boundaries or in between, misalignment of the masks, etc. In case large defects are found the foils will be discarded. All findings including their positions will be entered into the database.

Next the foils will undergo a High Voltage cleaning procedure in connection with a short term leakage current test of 20~min. The foils will be instantly connected to 500~V under nitrogen gas flow with an absolute humidity of 6000 ppmV or less.
The leakage currents between each segment and the ground electrode are recorded. The test system, see Fig.\,\ref{leakagecurrent}(a), will be described in more detail when discussing the advanced QA that uses an identical device. In addition and prior to the actual leakage current test the protection resistors between the High Voltage (HV) bus and the segments are measured to be within 7\% of the nominal value of 5~M$\Omega$. GEMs that do not pass the leakage current test, have a short circuit or resistor values out of specs are directly repaired at CERN PCB workshop.  

\subsection{Advanced QA}
\noindent The more time-consuming advanced tests at dedicated QA centers consist of a High Definition (HD) optical scanning, a long term leakage current test and gain-mapping of the GEM foils. The gain measurement is expensive in time and resources and will only be done for selected foils. The final goal is to correlate the results of the optical scanning with the gain uniformity measurements to predict the gain uniformity by the optical scans.  

\subsubsection{High definition optical scanning}
\noindent The HD optical scanning system for the GEM QA consists of a xyz-robot mounted on a glass table~\cite{mpgd15}. A camera optics system is attached to the z-axis. This system was designed and built at the Helsinki Institute of Physics. Technical details can be found in Tab.\,\ref{HDscanspecs}.

\begin{wraptable}[14]{r}{0.4\textwidth}\centering
\vspace{-0.1cm}\caption{HD scanner specs.\label{HDscanspecs}}
\scriptsize{\begin{tabular}{llr}\toprule
Device & Part & Details\\ \midrule
xyz-robot & table size  & 60$\times$120~cm\\
& xyz step size & 0.1 / 0.1 / 0.25$\mu$m\\\midrule
Camera & Type & Monochrome\\
& Sensor & CMOS / 1/2''\\
& Pixels (H x V) & 2560$\times$1920 \\  
& Pixel size & 2.2~$\mu$m\\ \midrule
Optics & Type & Telecentric\\ 
& Magnification & 0.5$\times$ \\ \midrule
Lighting & Coaxial Inline & LED (white)\\
& Ringlight & LED (white)\\
& Background & LED strip (white)\\ 
\bottomrule
\end{tabular}}
\end{wraptable}
The scanner is controlled via custom made software that allows semi-automated operation.

The GEM foils are scanned with a so-called two-exposure mode. Each image of the size 11.2$\times$8.4~mm is taken twice, one with the back-light and one with the ring- and coaxial inline- lights on. The two monochrome images are composed into a single image separated into different color channels and stored in a NAS connected to the scanner via Gigabit Ethernet link. The back-light exposure is used for the reconstruction of the inner polyimide holes and the foreground-light exposure for the copper holes. The largest foil -- OROC3 -- for example consists of $\sim$3600 single images. A whole scanning process can take up to 2 hours depending on the size of the GEM foil.

The reconstruction and analysis of the images and the classification of the found objects, whether it is a good hole or any kind of defect, is done via custom made software introduced in~\cite{timoerik}. However, the software has slightly changed over the years, especially the reconstruction of the copper holes. The polyimide holes are still found using the Canny algorithm. For the holes in the metal surface a convolutional network is used to determine whether a image pixel is part of the copper surface or the polyimide rim. Knowing the copper-edge pixels the hole contours can be found. From this point on the analysis continues as described in~\cite{timoerik}

\begin{figure*}[t!]
    \centering
    \begin{subfigure}[t]{0.98\textwidth}
        \centering
        \includegraphics[width=0.6\textwidth]{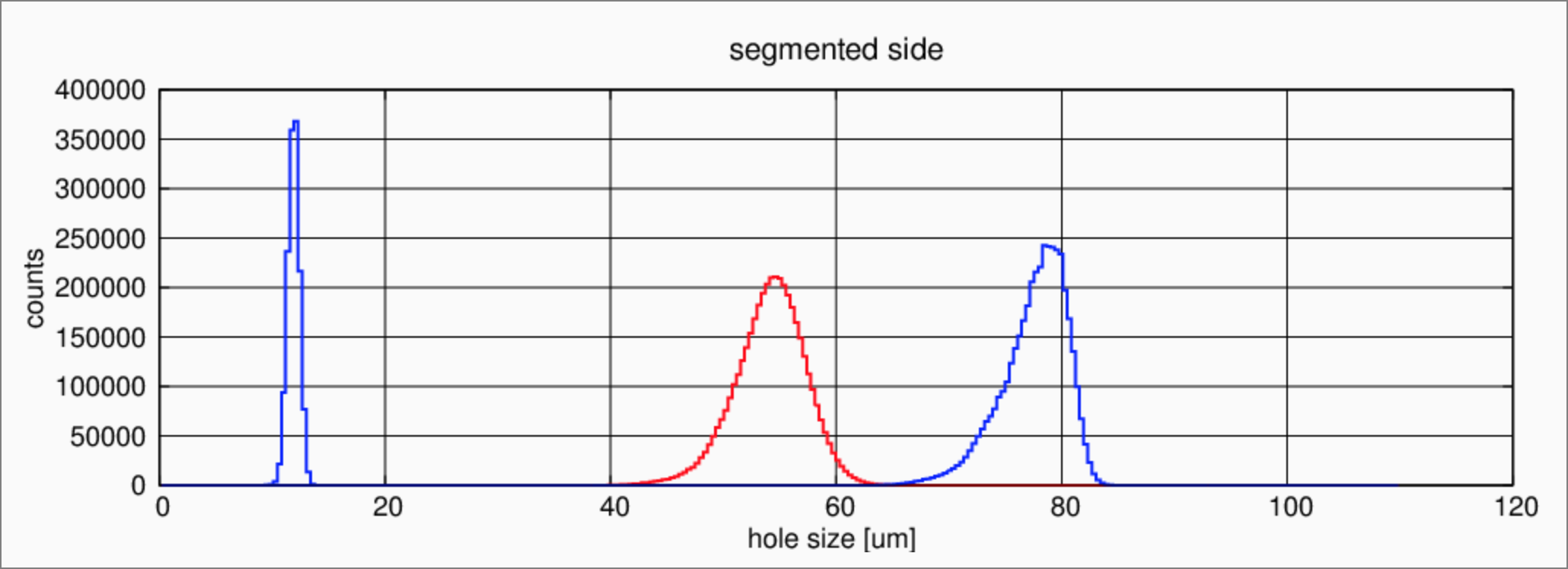}
        \caption{}
    \end{subfigure}%
    
    \begin{subfigure}[t]{0.49\textwidth}
        \centering
        \includegraphics[width=0.98\textwidth]{o3g3005_s_inner2d_m.png}
        \caption{}
    \end{subfigure}%
     ~
    \begin{subfigure}[t]{0.49\textwidth}
        \centering
        \includegraphics[width=0.98\textwidth]{o3g3005_s_outer2d_m.png}
        \caption{}
    \end{subfigure}
    \caption{(a) 1 d histogram of the rim sizes and the polyimide and copper hole diameters of the segmented side of the GEM foil O3\_G3\_005. 2-d maps of (b) the polyimide hole diameters and (c) the copper hole diameters of the same GEM foil.\label{hdplots}}
\end{figure*}

The reconstructed results contains among others position and size information of every single GEM hole and recognized defects. 
Maps and histograms are generated from the output. Example plots can be seen in Fig.\,\ref{hdplots}.
The scanning results are used to predict the performance of the individual GEM foils as shown in~\cite{timoerik,timophilly}. In addition the defects are classified using a neural network.

\subsubsection{Long term leakage current test}
\begin{figure*}[t!]
    \centering
    \begin{subfigure}[t]{0.3\textwidth}
        \centering
        \includegraphics[height=3.5cm]{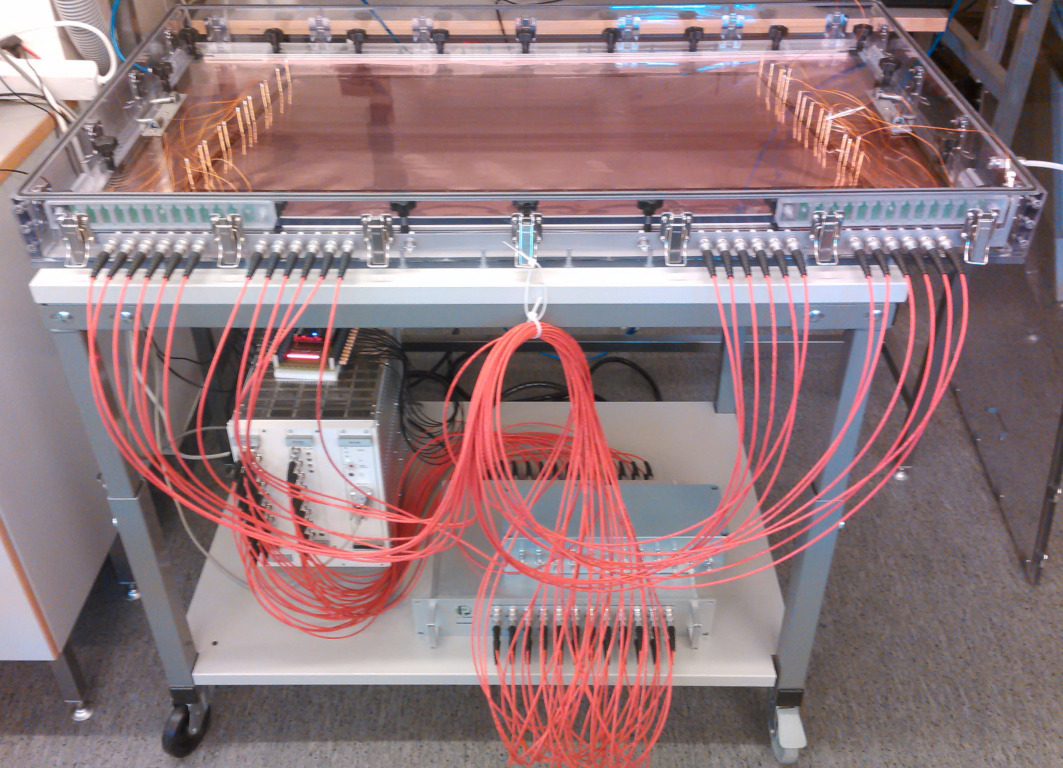}
        \caption{}
    \end{subfigure}%
    ~ 
    \begin{subfigure}[t]{0.7\textwidth}
        \centering
        \includegraphics[height=3.5cm]{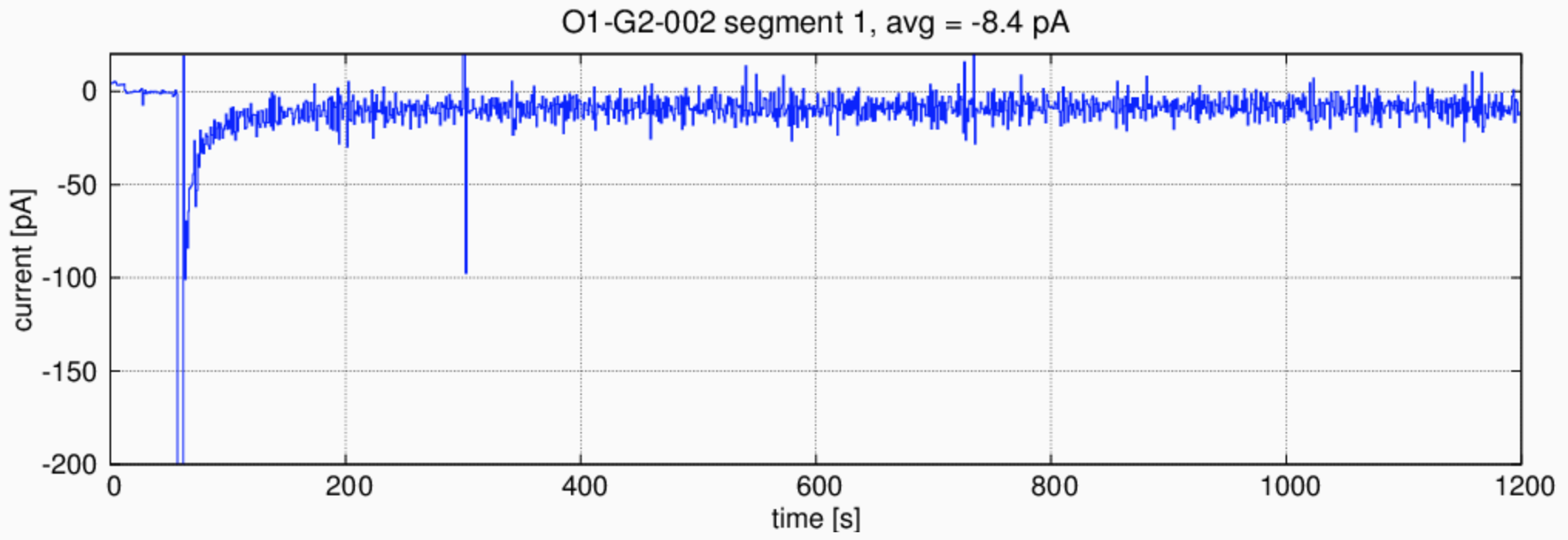}
        \caption{}
    \end{subfigure}
    \caption{(a) The leakage current test system and (b) an example plot of a measured leakage current of one segment of a good GEM foil.\label{leakagecurrent}}
\end{figure*}

\noindent The test system for short and long term leakage currents consists of a HV power supply, a picoammeter and a closed gas tight box hosting the GEM to be tested (see Fig.\,\ref{leakagecurrent}(a)). The HV power supplies vary between QA centers but the important specifications are similar, output voltage $\geq$ 500~V, output current $\leq$ 1~mA and ripple noise $\leq$10~mV. 

The TPC ROC GEM foils are divided into 18 to 24 HV segments of roughly equal area. The leakage currents are measured by a Picologic 24 channel picoammeter~\cite{zagreb}.
The foils are tested in custom made acrylic boxes with separate inlays for all 4 GEM foil sizes. The box is flushed thoroughly with nitrogen until the humidity inside is less then 10\%. The GEM foils are equipped with protection resistors of 5~M$\Omega$. This made it necessary to produce the inlays with very high precision to take care that the special spring connectors with conical tips touch the resistors at the solder point of the GEM segment for good electric contact. As all GEM segments are connected to the HV bus via these resistors, the measured currents are not independent of each other. But it has been shown that it is possible to reliably measure the leakage currents of each segment in case the foils do not show shortcuts in any segment. The additional parallel circuit due to the resistors to the HV bus introduces lower practical limits of 160~pA compared to the true current limit criteria of 500~pA. 

The system is operated via custom software that allows semi-automation. The voltage is first turned on to 500~V before connecting the outlet to the GEM foil. This is due to the expected cleaning effect of instantly turned on voltage in case dust exists in the vicinity of the GEM holes. For the long term test foils are measured for at least 5~hours. In case a measurement is started in the evening it is continued until the next morning (12~h).

Data is saved for further analysis in a NAS. A common data format is used that includes  date and time information, the currents from all segments and the power supply voltage and current. After the measurements the data files are uploaded to the database that allows direct plotting and analysis of the leakage currents. In Fig.\,\ref{leakagecurrent}(b) an example plot is shown of a leakage current measurement generated with a feature of the database.

\subsubsection{Gain mapping}
\begin{figure*}[t!]
    \centering
    \begin{subfigure}[t]{0.5\textwidth}
        \centering
        \includegraphics[height=4.5cm]{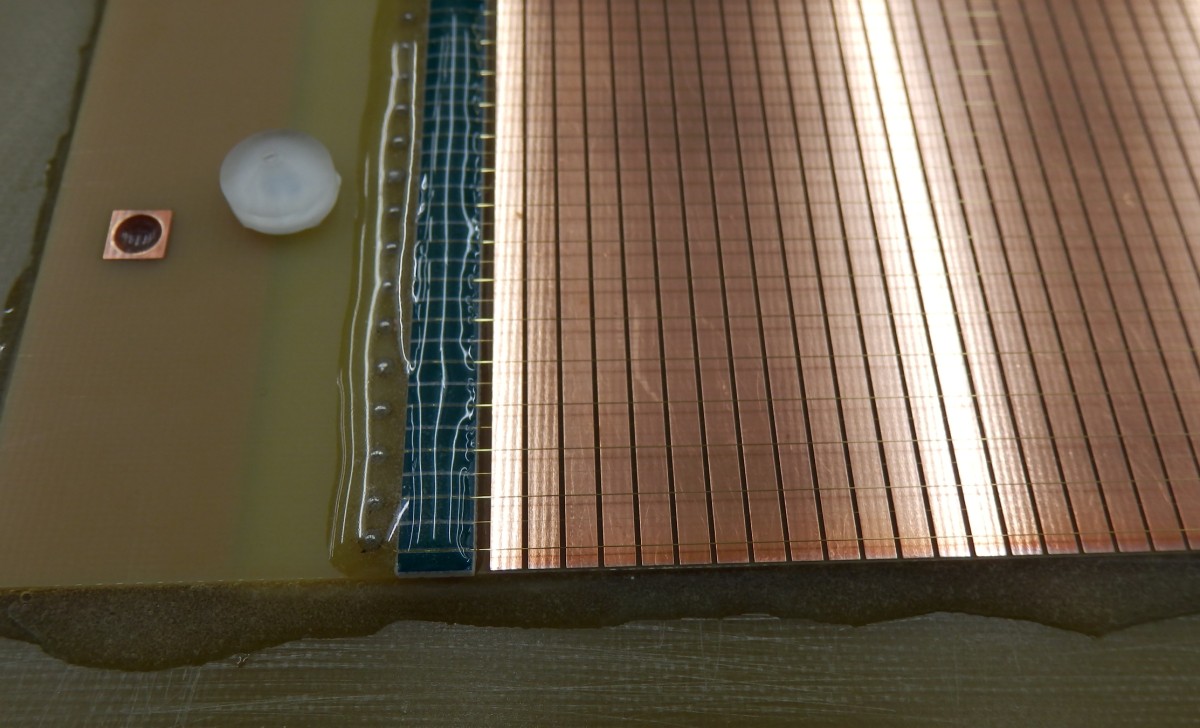}
        \caption{}
    \end{subfigure}%
    ~ 
    \begin{subfigure}[t]{0.5\textwidth}
        \centering
        \includegraphics[height=4.5cm]{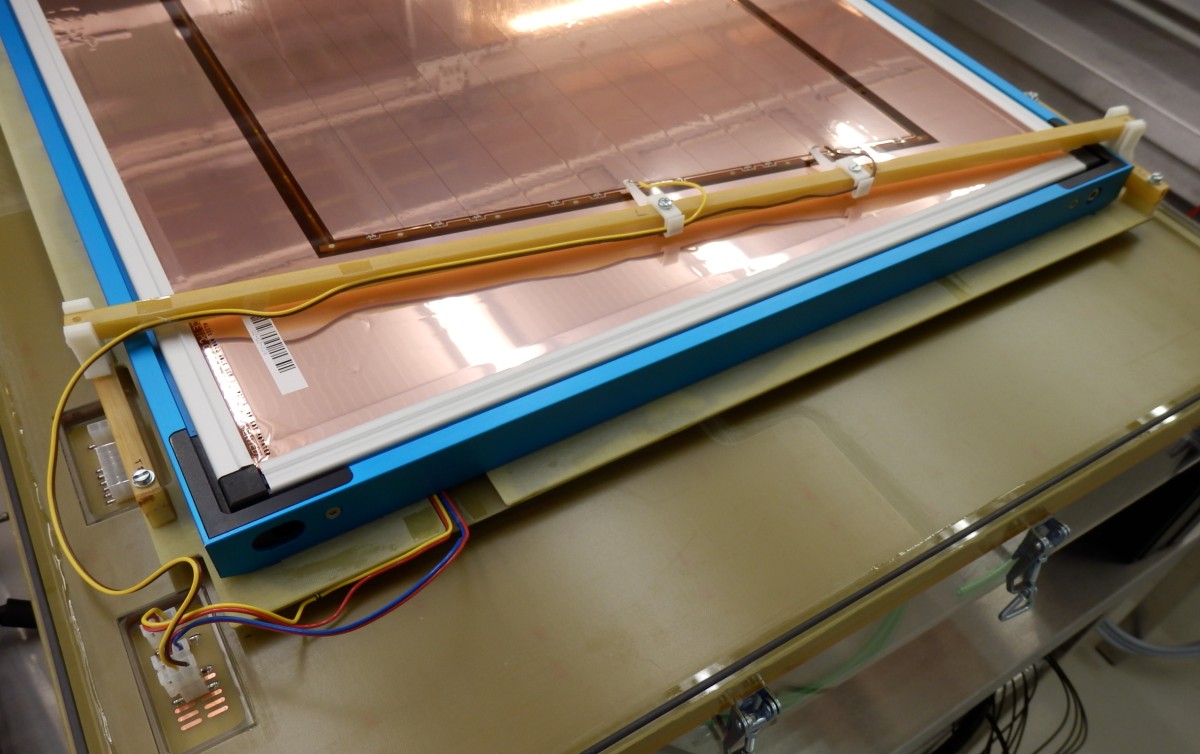}
        \caption{}
    \end{subfigure}
    \caption{(a) The underlying MWPC of the GEM gain test system and (b) a GEM foil mounted on top of it.\label{GMTS}}
\end{figure*}

\noindent As earlier introduced and mentioned in~\cite{mpgd15} a full size gain mapping device has been built in the Budapest QA center that fits the largest of the GEMs. A MultiWire Proportional Chambers (MWPC) placed on top of a custom built readout are used to measure the gain of the GEMs. Fig.\,\ref{GMTS}(a) shows the underlying mutliwire proportional chamber and Fig.\,\ref{GMTS}(b) a mounted GEM foil inside the test system. The foil is irradiated with an $^{55}$Fe source and the environmental changes are monitored throughout the measurement. However, it is not possible to map the gain of every single foil as there is not enough time and resources. Strict measurement conditions for accuracy and repeatability are enforced. Making a single measurement takes more than 24~h. The relative gain measurement accuracy is 5\% with resolution of 4~mm $\times$ 3~mm.
Because not all GEMs can be tested it is foreseen to predict the gain based on the results of the HD optical scanning~\cite{timoerik}.

\subsection{Criteria}
\noindent At present time fine tuning of the QA criteria is still ongoing. 
Tab.\,\ref{criteria} shows the present criteria. 
\begin{table}[!htb]\centering
\caption{QA criteria.\label{criteria}}
\scriptsize{\begin{tabular}{lc|cccc|c|c}\toprule
 \multirow{2}{*}{Criteria} & \multirow{2}{*}{Basic QA\footnotemark} & \multicolumn{4}{c|}{Optical QA} & Long-term & Gain \\ \cmidrule{3-6}
&& \multicolumn{3}{c|}{hole size based} & map based & stability QA & scanning  QA\\ \midrule
\multirow{2}{*}{Color code}&\multirow{2}{*}{I$_\mathrm{leak}$}& inner/outer &\multirow{2}{*}{rim mean} &inner/outer & \multirow{2}{*}{ptp} &\multirow{2}{*}{I$_\mathrm{leak}$}&\multirow{2}{*}{uniformity}\\
&&RMS&&deviation&&&\\
\midrule
\cellcolor{green}Green& <  500  pA & <  4 $\mu$m & <  15 $\mu$m & <  5 $\mu$m & <  5 $\mu$m & <  500  pA & <  10\%\\ 
\cellcolor{yellow}Yellow& n/a & n/a & 15-19 $\mu$m & 5-10 $\mu$m & 5-10 $\mu$m & < 500 pA / non stable & > 10\% \\ 
\cellcolor{orange}Orange & n/a & >  4  $\mu$m & >  19  $\mu$m & >  10  $\mu$m & >  10  $\mu$m & n/a & n/a\\
\cellcolor{red}Red& >  500  pA & n/a & n/a & n/a & n/a & >  500  pA &n/a\\
\bottomrule
\end{tabular}}
\end{table}
\footnotetext{Table not complete; missing criteria a.o. for coarse optical inspection, protection resistors, sparks at 500~V, etc.}
A classification of the GEMs is done using a traffic light system. The GEM color is based on the worst criteria that is met. Red foils are discarded. Green and yellow foils are used for the ROC production. The orange class is solely for foils that did not pass the advanced optical QA criteria but are otherwise green or yellow and show stable electrical behavior. In case no green or yellow foils are available, orange foils will be used.

It is of special importance to investigate the performance of GEM stacks including orange GEM foils. At the current stage it is not clear if the chamber performance can be substantially improved by removing GEM foils of orange classification.  

\section{Workflow}
\noindent In average 40 GEM foils per month are received by the CERN Basic QA center from the GEM production workshop. The GEMs are cut from the production frames and placed into special aluminum profile frames for stretching purposes. 
After the basic QA the foils are carefully packed into dedicated envelopes, mounted into a GEM Transport System (GTS) and sent in equal amounts to the advanced QA centers in Budapest or Helsinki.

The GEMs are unpacked in the QA centers and put in vertical hanging position into dry cabinets for storage. First the foils are tested with the HD scanner. After that the long term leakage current measurement is done. In case the leakage current test shows undesired activity in terms of high currents or short circuits, the particular foil is one time locally cleaned with electrostatic silicon rollers. If this does not recover the foil it will be sent back to CERN for re-cleaning. Good foils are put back into the GTS and sent to the framing institutes.  

The QA data is then uploaded to the common database in ASCII format. 
The full history of each individual GEM foil can be found in the database. The database has features to directly plot on demand graphs and histograms from the uploaded data such as leakage currents per segment, 1-d histograms and maps of the hole sizes and defect maps. The database is not only serving the GEM QA but includes information related to the ROC production, infrastructure and logistics. 

\section{Summary}
\noindent Over the past years a protocol for the quality assurance of the ALICE TPC ROC GEM foils was established.
Since January 2017 -- the beginning of the ROC production -- foils have been delivered constantly by the GEM production workshop. The QA centers started successfully and in time. Currently approximately 40 foils per month are tested and classified (Basic QA 40 foils, advanced QA 20 foils in each QA center). With the present QA criteria the yield of foils that pass is close to 90\%.


\end{document}